\documentclass[aps,showpacs]{revtex4}
\usepackage{graphicx}
\begin{document}

\title{Vaccum solutions of five-dimensional Einstein equations generated by
\\inverse scattering method II : Production of black ring solution}

\author{Shinya Tomizawa$^1$\footnote{tomizawa@sci.osaka-cu.ac.jp}
and Masato Nozawa$^2$\footnote{nozawa@gravity.phys.waseda.ac.jp}}

\affiliation{ Department of Mathematics and Physics,
Graduate School of Science, ${}^1$Osaka City University,
3-3-138 Sugimoto, Sumiyoshi, Osaka 152-8551, Japan\\
${}^2$Department of Physics, Waseda University, 3-4-1,
Okubo, Shinjyuku-ku, Tokyo, 169-8555, Japan}

\date{\today}

\begin{abstract}
We study vacuum solutions of five-dimensional Einstein equations
generated by the inverse scattering method.
We reproduce the black ring solution which was found by Emparan
and Reall by taking the Euclidean Levi-Civita metric plus 
one-dimensional flat space as a seed.
This transformation consists of two successive processes;
the first step is to perform the three-solitonic transformation
of the Euclidean Levi-Civita metric with one-dimensional flat space
as a seed. The resulting metric is the Euclidean C-metric 
with extra one-dimensional flat space.
The second is to perform the two-solitonic transformation
by taking it as a new seed.
Our result may serve as a stepping stone to find
new exact solutions in higher dimensions.
\end{abstract}


\maketitle

\section{Introduction}
In recent years, studies of black holes in higher dimensions
have attracted much attention in the context of string theory
and the brane world scenario.
In fact, it has been predicted that higher-dimensional black holes
would be produced in a future linear collider
\cite{BHinCollider,kanti}.
Such physical phenomena are expected not only to give us
a piece of evidence for the existence of extra dimensions
but also to help us to draw some information toward quantum gravity.
Studies on the classical equilibrium state of black holes are important
since it is extrapolated that we may detect the Hawking evaporation
after the formation of stationary black holes in a collider.

Some of classical studies on higher-dimensional black holes
show that they have much more complicated and richer structure
than four-dimensional ones.
For instance, the topology of the event horizon
in higher dimensions cannot be uniquely determined
\cite{Cai,Helfgott,galloway}
in contrast to four-dimensional ones, which is restricted only to
the two sphere under null energy condition \cite{Hawking,hawking_ellis}.
In five dimensions, however, the possible geometric types of the horizon topology
are either $S^3$ or $S^1\times S^2$ \cite{Cai},
and in dimensions higher than five, more complicated
\cite{Helfgott,galloway}.

As for the asymptotically flat, static solutions
of higher-dimensional vacuum Einstein equations, the Schwarzschild-Tangherlini
solution \cite{T-shwarzschild} is the unique solution \cite{shiromizu},
and moreover, which is stable against gravitational perturbations
\cite{kodama}.
These are common properties as the counterpart in four dimensions. However, the situation radically changes for the stationary spacetime;
for asymptotically flat, stationary and axisymmetric solutions
of the five-dimensional vacuum Einstein equations,
Emparan and Reall discovered the black ring solution
whose horizon topology is $S^1\times S^2$ \cite{Emparan:2001wn}.
The black ring is rotating along $S^1$ direction,
which is necessary for the ring to keep a balance
against its self-gravitational attractive force.
The black ring is regular everywhere and
has no closed timelike curves on and outside the event horizon.
In addition to the black ring solution,
the rotating black hole solution with $S^3$ horizon topology
had been already found by Myers and Perry \cite{Myers:1986un}.
This means that, in the sense of four dimensions,
uniqueness theorems of black hole does not
hold in higher dimensions \cite{uniqueness},
although it has been shown that the five-dimensional
Myers-Perry solution is unique if the topology is restricted to $S^3$
and the spacetime admits three commuting Killing vectors
\cite{Morisawa:2004tc}.
Furthermore, as several arguments suggest that
the black ring solution is unstable \cite{nozawa,arcioni,Gregory},
there might exist a new family of solutions with less symmetry
as the final phase of instability \cite{reall}.
These rich structures in higher-dimensional spacetimes
have made it desirable to use techniques
to find a new solution.

A new stationary and axisymmetric black ring solution with asymptotic
flatness, which is rotating only in two sphere direction,
was found by Mishima and Iguchi~\cite{Mishima:2005id,Figueras}
by using one of the solitonic solution-generating
techniques \cite{Castejon-Amenedo:1990b}.
Solitonic solution-generating methods are
mainly classified into two types.
One is called B\"acklund transformation \cite{Harrison,Neugebauer},
which is basically the technique to generate 
a new solution of the Ernst equation.
The other is the inverse scattering technique,
which Belinski and Zakharov~\cite{Belinskii} developed
as an another type of soliton-generating technique.
Both methods have produced vacuum solutions from a certain known vacuum
solution and succeeded in generation of some
four-dimensional exact solutions.
As a matter of fact, 
the Kerr black hole solution, the multi Kerr black hole solutions
and the Tomimastu-Sato solutions can be obtained from the Minkowski seed
(other physically interesting solutions can also be obtained
by these methods \cite{solitonbook,exact}).
The latter technique is essentially based on the fact that the Einstein's
second-order nonlinear partial differential equations can be replaced  with
a pair of first-order linear partial differential
equations called Lax pair.
It can be immediately generalized to $D$-dimensional vacuum spacetimes
with $(D-2)$-commuting Killing vectors. The complete integrability in five dimensional Einstein-Maxwell systems is discussed in Ref.~\cite{Yazadjiev}. 
As discussed in \cite{Tomizawa},
the former cannot generate black hole/ring solutions
with two angular momentum components due to its ansatz,
the latter has an advantage to be able to produce such solutions apart from
whether these solutions are regular or not, 
though this happens only in five dimensions.
Therefore, there would be the possibility that
new black hole/ring solutions can be found by using this method.

Recently, some of higher-dimensional black hole/ring solutions
have been generated by means of the inverse scattering method.
As an infinite number of static solutions of the five-dimensional
vacuum Einstein equations with axial symmetry,
the five-dimensional Schwarzschild solution
and the static black ring solution were reproduced \cite{Koikawa},
which gave the first example of the generation of 
a higher-dimensional asymptotically flat black hole solution 
by the inverse scattering method.
The Myers-Perry solution with single and double angular momenta
were regenerated from the Minkowski \cite{Tomizawa,Azuma} and
some technical seed~\cite{Pomeransky:2005sj}, respectively.
The black ring solutions with $S^2$ rotation 
was also generated by using this method
from the Minkowski seed~\cite{Tomizawa}.
Albeit several articles have been devoted to construct
the black ring solution with $S^1$ rotation by this method,
it has not been successful so far.
In this article, we reproduce the $S^1$-rotating black ring  solution
from the Levi-Civita solution via the inverse scattering method.

This article is organized as follows.
In Sec. \ref{sec:preliminary},
the outline of the inverse scattering
technique is summarized.
In Sec. \ref{sec:seed}, we start from the Euclidean
Levi-Civita metric plus one-dimensional flat space,
the C-metric plus one-dimensional flat space is obtained
as a three-soliton solution. By taking this solution as a new seed,
the black ring solution is recaptured as a two-soliton solution.
We will see this is indeed the case in Sec. \ref{sec:black ring}.
We will frame our conclusion in Sec. \ref{sec:conclusion}.

\section{Preliminary}
\label{sec:preliminary}

We will give a summary account of the inverse scattering method
developed by Belinski and Zakharov \cite{Belinskii},
which is straightforwardly applied into five dimensions.

\subsection{Five-dimensional $n$-soliton solutions}
\label{sec:general}

We consider the asymptotically flat, five-dimensional
stationary and axisymmetric vacuum spacetime
with three commuting Killing vector fields $V_{(i)}$
($i=1,2,3$) following the argument in
\cite{weyl,Harmark:2004rm}.
The commutativity of Killing vectors $[V_{(i)}, V_{(j)}]=0$
enables us to find a coordinate system such that
$V_{(i)}=\partial/\partial x^i~(i=1, 2, 3)$ and
the metric is independent of the coordinates $x^i$,
where $(\partial/\partial x^1)$ is the Killing vector field
associated with time translation and
$(\partial/\partial x^2),(\partial/\partial x^{3})$
denote the spacelike Killing vector fields with closed orbits.
In the following section, we  will put
$x^1=t,\ x^2=\phi,$ and $x^3=\psi$.
Here we invoke the theorem in
\cite{weyl,Harmark:2004rm}:

{\it In D-dimensional spacetime, let
$V_{(i)}, i=1,...,D-2$, be $(D-2)$-commuting
Killing vector fields such that
\begin{enumerate}
\item
$V_{(1)}^{[\mu _1}V_{(2)}^{\mu _2}\cdots
V_{(D-2)}^{\mu _{D-2}]}D^\nu V_{(i)}^{\rho }=0$
holds at at least one point of the spacetime
for a given $i=1, ..., D-2$.

\item
$V_{(i)}^\nu R_\nu ^{[\rho}
V_{(1)}^{\mu _1}V_{(2)}^{\mu _2}\cdots
V_{(D-2)}^{\mu _{D-2}]}=0 $ holds for all
$i=1, ..., D-2.$
\end{enumerate}
Then the two-planes
orthogonal to the Killing vector fields
$V_{(i)}, i=1, ..., D-2,$ are integrable.}

Thanks to the axisymmetry,
condition 1. holds on the axis of rotation.
Meanwhile, condition 2. is automatically satisfied
as long as we restrict
ourselves to the vacuum solutions of Einstein equations
$R_{\mu \nu }=0$, where $\mu ,\nu $ run over all spacetime indices.
In such a spacetime, the metric can be written
in the canonical form~\cite{weyl,Harmark:2004rm} as
\begin{eqnarray}
ds^2=f(d\rho^2+dz^2)+g_{ij}dx^idx^j,
\label{eq:canonical}
\end{eqnarray}
where $f=f(\rho,z)$ and $g_{ij}=g_{ij}(\rho,z)$
are a function and an induced metric on the
three-dimensional space, respectively.
Both of them depend only on the coordinates $\rho$ and $z$.
Here it is the most convenient to choose
the $3\times 3$ matrix $g=(g)_{ij}$ as to satisfy the condition
\begin{eqnarray}
{\rm det}g=-\rho^2.
\label{eq:det}
\end{eqnarray}
This is compatible with the vacuum Einstein equations
$g^{ij}R_{ij}=0$, which reduces to
$(\partial _\rho ^2+\partial _z^2)(-{\rm det}g)^{1/2}=0$.
It follows from $R_{ij}=0$ that the matrix
$g$ satisfies the solitonic equation
\begin{eqnarray}
(\rho g_{,\rho} g^{-1})_{,\rho}+(\rho g_{,z}g^{-1})_{,z}=0.
\label{eq:soliton}
\end{eqnarray}
From the other components of the Einstein
equations $R_{\rho\rho}-R_{zz}=0$ and $R_{\rho z}=0$,
we obtain the equations which determine the function $f(\rho,z)$
for a given solution of the solitonic equation (\ref{eq:soliton})
\begin{eqnarray}
(\ln f)_{,\rho}&=&-\frac{1}{\rho}+\frac{1}{4\rho}{\rm Tr}(U^2-V^2),
\label{eq:f1} \\
(\ln f)_{,z}&=&\frac{1}{2\rho}{\rm Tr}(UV),\label{eq:f2}
\end{eqnarray}
where the $3\times 3$ matrices $U(\rho,z)$ and $V(\rho,z)$
are defined by
\begin{eqnarray}
U:=\rho g_{,\rho}g^{-1}, \qquad V:=\rho g_{,z}g^{-1}.
\end{eqnarray}
The integrability condition with respect to $f$ is automatically
satisfied for the solution $g$ of Eq.~(\ref{eq:soliton}).
Note also that $R_{\rho \rho }+R_{zz}=0$ is consistent with
the solution (\ref{eq:soliton}), (\ref{eq:f1}) and (\ref{eq:f2}),
i.e., this system is not overdetermined.

Although our immediate goal is to solve the differential
equations (\ref{eq:soliton}), it cannot be generally solved
due to its non-linearity.
But in analogy with the soliton technique,
we can find the Lax pair for the matrix
equations (\ref{eq:soliton}).
We consider Schr\"odinger type equations
for the $3\times 3$ matrix $\Psi (\lambda ,\rho ,z)$
as in four dimensions \cite{Belinskii,solitonbook};

\begin{eqnarray}
D_{1}\Psi =\frac{\rho V-\lambda U}
{\lambda ^2+\rho ^2}\Psi,
\qquad
D_{2}\Psi =\frac{\rho U+\lambda V}
{\lambda ^2+\rho ^2}\Psi,
\label{eq:Laxpair}
\end{eqnarray}
where $\lambda $ is a complex spectral
parameter independent of $\rho $ and $z$.
The differential operators
$D_1$ and $D_{2}$ are defined as
\begin{eqnarray}
D_{1}:=\partial _z-\frac{2\lambda ^2}
{\lambda ^2+\rho ^2}\partial _\lambda ,
\qquad
D_{2}:=\partial _\rho +\frac{2\lambda \rho }
{\lambda ^2+\rho ^2}\partial _\lambda ,
\label{eq:LAeq}
\end{eqnarray}
which can be shown to commute $[D_1, D_2]=0$.
Note that Eq. (\ref{eq:LAeq}) is invariant under
the transformation $\lambda \to -\rho ^2/\lambda $.
Then the compatibility condition
$[D_1, D_2]\Psi =0$ reduces to
the Einstein equations (\ref{eq:soliton}) with
\begin{eqnarray}
g(\rho ,z)=\Psi (0, \rho, z).
\label{eq:initial}
\end{eqnarray}
It deserves to note that the Einstein's second-order
nonlinear partial differential
equations (\ref{eq:soliton}) are reduced to
a pair of first-order linear partial differential equations
(\ref{eq:Laxpair}).

Let $g_0, U_0, V_0$ and $\Psi_0$ be
particular solutions of Eq. (\ref{eq:soliton})
and (\ref{eq:Laxpair}).
We shall call the known solution $g_0$
the seed solution.
We are going to seek a new solution
of the form
\begin{eqnarray}
\Psi =\chi \Psi_0,
\label{eq:DM}
\end{eqnarray}
which leads the following equations 
that the dressing matrix
$\chi (\lambda , \rho ,z)$ must satisfy 
\begin{eqnarray}
D_1 \chi =\frac{\rho V-\lambda U}
{\lambda ^2+\rho ^2}\chi
-\chi \frac{\rho V_0-\lambda U_0}
{\lambda ^2+\rho ^2},\nonumber \\
\label{eq:DMeq}\\
D_2 \chi =\frac{\rho U+\lambda V}
{\lambda ^2+\rho ^2}\chi
-\chi \frac{\rho U_0+\lambda V_0}
{\lambda ^2+\rho ^2}.\nonumber
\end{eqnarray}

In order for the solutions $g(\rho ,z)$
to be real and symmetric,
we impose the following conditions
on the dressing matrix $\chi$,
\begin{eqnarray}
\bar\chi (\bar \lambda ,\rho ,z)=\chi (\lambda ,\rho ,z),
\qquad
\bar\Psi (\bar \lambda ,\rho ,z)=\Psi (\lambda ,\rho ,z),
\label{eq:realg}
\end{eqnarray}
and
\begin{eqnarray}
g=\chi (-\rho ^2/\lambda ,\rho ,z)g_0
{}^T\hspace{-0.1cm}\chi (\lambda ,\rho ,z),
\label{eq:symg}
\end{eqnarray}
where $\bar \chi $ and ${}^T\chi$  denote
complex conjugation and the transposition of $\chi$.
From Eqs. (\ref{eq:DM})
and (\ref{eq:symg}), the dressing matrix
$\chi$ asymptotes to a unit matrix
$\chi \to I$ as $ \lambda \to \infty $.

The general $n$-soliton solutions for the
matrix $g$ are generated due to the presence of the
simple poles of the dressing matrix on the complex
$\lambda $-plane:
\begin{eqnarray}
\chi =I+\sum_{k=1}^n\frac{S_k}
{\lambda -\mu _k},
\label{eq:DMform}
\end{eqnarray}
where the matrices $S_k$ and the position
of the pole $\mu _k$ depend only on
the variables $\rho $ and $z$.
Here and hereafter, the subscript $k, l$
counts the number of solitons.
It is the characteristic feature of solitons 
that the dressing matrix 
$\chi$ is represented as the meromorphic function
on the complex $\lambda$-plane.
Pole trajectories $\mu _k (\rho ,z)$
are determined by
the condition that the left-hand side of Eq.
(\ref{eq:DMeq}) have no poles of second-order
at $\lambda =\mu _k$, which yields following two
differential equations for $\mu _k (\rho ,z)$:
\begin{eqnarray}
\mu _{k,z}=-\frac{2\mu _k^2}
{\mu _k^2+\rho ^2},
\qquad
\mu _{k,\rho }=\frac{2\rho \mu _k}
{\mu _k^2+\rho ^2},
\label{eq:eqmuk}
\end{eqnarray}
which are expressed by the solutions of the
following quadratic equations
\begin{eqnarray}
\mu _k^2+2(z-w_k)\mu _k-\rho ^2=0,
\label{eq:quadmu}
\end{eqnarray}
where $w_k$ are arbitrary constants.
Solving Eq. (\ref{eq:quadmu}), one can easily find
\begin{eqnarray}
\mu _k=w_k-z\pm \sqrt{(z-w_k)^2+\rho ^2}.
\label{eq:mup}
\end{eqnarray}
Since the matrices $S_k$ are degenerate
at the poles $S_k \chi^{-1}(\mu _k)=0$,
which follows from the condition $\chi \chi^{-1}=I$
at $\lambda =\mu _k$, it is possible to
write down the matrix elements of $S_k$
in the form
\begin{eqnarray}
(S_k)_{ij}=n_i^{(k)}m_j^{(k)}.
\end{eqnarray}
The fact that Eq. (\ref{eq:DMeq}) has no
residues at the poles $\lambda =\mu _k $ leads
to obtain the vectors $m_i^{(k)}$ as
\begin{eqnarray}
m^{i(k)}= m_{0j}^{(k)}[\Psi_0^{-1}
(\mu_k, \rho ,z)]^{ji},
\label{eq:vectorm}
\end{eqnarray}
where $m_{0i}^{(k)}$ are arbitrary constants.
The vectors $n_i^{(k)}$, on the other hand,
are determined
by the condition that Eq. (\ref{eq:symg})
is regular at $\lambda =\mu _k$ as
\begin{eqnarray}
n_i^{(k)}=\sum _{l=1}^n\mu _k^{-1}(\Gamma ^{-1})
_{kl}L_{i}^{(l)},
\label{eq:vectorn}
\end{eqnarray}
where the vectors $L_i^{(k)}$ and
the symmetric matrix $\Gamma _{kl}$ are given by
\begin{eqnarray}
L_i^{(k)}&=&m^{j(k)}(g_0)_{ij},
\label{eq:vectorl}\\
\Gamma _{kl}&=&\frac{m^{i(k)}(g_0)_{ij}m^{j(l)}}
{\rho ^2+\mu _k\mu _l},
\label{eq:gamma}
\end{eqnarray}
respectively. Therefore one can now find
from Eq. (\ref{eq:initial}),
(\ref{eq:DM}) and (\ref{eq:DMform}) that
the matrix $g$ becomes
\begin{eqnarray}
g^{({\rm unphys})}_{ij}
&=&\Psi (0, \rho ,z)_{ij} \\
\nonumber \\
&=&(g_0)_{ij}-\sum_{k,l=1}^n(\Gamma^{-1})_{kl}\mu _k
^{-1}\mu _l^{-1}L_i^{(k)}L_j^{(l)}.
\label{eq:unphys1}
\end{eqnarray}
This metric does not necessarily meet the condition
${\rm det}g=-\rho^2$, which we have denoted
$g^{({\rm unphys})}$.
In order to satisfy the gauge condition
${\rm det}g=-\rho^2$,
the metric should be appropriately normalized.
One example is to normalize
all the metric components by the same weight as

\begin{eqnarray}
g^{{\rm (phys)}}=(-1)^{{n}/{3}}\rho^{-{2n}/{3}}
\left(\prod_{k=1}^n\mu_k^{{2}/{3}}\right)g^{{\rm (unphys)}},
\label{eq:nor}
\end{eqnarray}
where $g^{{\rm (phys)}}$ is the metric
which fulfills the condition ${\rm det}g=-\rho^2$.
Actually, the four-dimensional Kerr solution is obtained
similarly by the overall normalization as Eq. $(\ref{eq:nor})$.
Substituting the physical metric solution
$g^{{\rm (phys)}}$ given by Eq. (\ref{eq:nor})
into Eq. (\ref{eq:f1}) and  (\ref{eq:f2}),
we obtain a physical value of $f$ as

\begin{eqnarray}
f=C_0f_0\rho^{-{n(n-1)}/{3}}{\rm det}
(\Gamma_{kl})\prod_{k=1}^n\left[\mu_k^{{2(n+2)}/{3}}
(\mu_k^2+\rho^2)^{{-1}/{3}}\right]\cdot
\prod_{k>l}^n(\mu_k-\mu_l)^{{-4}/{3}},
\end{eqnarray}
where $C_0$ is an arbitrary constant, and $f_0$ is a value of $f$
corresponding to the seed $g_0$.

\subsection{Relation between two diagonal generating matrices}
\label{sec:rescale}

In this subsection, we study how the diagonal generating 
matrices are mutually ralated. 
Let $\Psi ={\rm diag}(\psi_1,\psi_2,\psi_3)$ and  
$\tilde \Psi={\rm diag}(\tilde \psi_1,\tilde \psi_2,\tilde \psi_3)$ 
be generating matrices
corresponding to the diagonal seeds 
$g={\rm diag}(g_1,g_2,g_3)$ and 
$\tilde g={\rm diag}(\tilde g_1,\tilde g_2,\tilde g_3)$
satisfying Eq. (\ref{eq:soliton}), respectively.
If they are related as 
$\tilde g_i=\Omega_i g_i (i=1, 2, 3)$ for 
$\Omega _i=\Omega _i(\rho ,z)$,
it follows from the Einstein equations (\ref{eq:soliton})
that the logarithm of $\Omega _i$ 
must be harmonic functions 
on the three-dimensional Euclidean space:
\begin{eqnarray}
\triangle \ln \Omega_i:=\Biggl(\frac{\partial^2}
{\partial\rho^2}+\frac{1}{\rho}
\frac{\partial}{\partial \rho}
+\frac{\partial^2}{\partial z^2} \Biggr)\ln  \Omega_i=0.
\label{eq:harmo}
\end{eqnarray}
Hence, the generating matrix
$\tilde \Psi={\rm diag}(\tilde\psi_1,\tilde\psi_2,\tilde\psi_3)$
corresponding to $\tilde g$ satisfies

\begin{eqnarray}
& &D_1 (\ln\tilde\psi_i)=\frac{\rho V_i-\lambda U_i}
{\rho^2+\lambda^2}+\frac{\rho^2(\ln \Omega_i)_{,z}-
\rho\lambda (\ln \Omega_i)_{,\rho}}{\rho^2+\lambda^2},\nonumber
\\
\label{eq:con1}
\\
& &D_2 (\ln\tilde\psi_i)=\frac{\rho U_i+\lambda V_i}
{\rho^2+\lambda^2}+\frac{\rho^2 (\ln \Omega_i)_{,\rho}
+\rho\lambda (\ln \Omega_i)_{,z}}{\rho^2+\lambda^2}.\nonumber\label{eq:con2}
\end{eqnarray}
where $U_i=\rho g_{i, \rho}g_i^{-1}$ and 
$V_i=\rho g_{i, z}g_i^{-1}$.

Since these equations are linear, we can easily find that
the solutions of Eq. (\ref{eq:con1}) %and (\ref{eq:con2})
are expressed in the form of
$\tilde \psi_i=\hat \psi (\Omega_i)\cdot\psi_i$,
where $\psi_i$ 
is the solution of Eq. (\ref{eq:Laxpair})
for the metric $g$, and  $\hat \psi(\Omega_i)$
is the solution of the following equations:
\begin{eqnarray}
D_1\left(\ln \hat \psi(\Omega_i)\right)
=\frac{\rho^2(\ln \Omega_i)_{,z}
-\rho\lambda (\ln \Omega_i)_{,\rho}}
{\rho^2+\lambda^2},\nonumber \\
\label{eq:con3}\\
D_2\left(\ln\hat \psi(\Omega_i)\right)
=\frac{\rho^2 (\ln \Omega_i)_{,\rho}
+\rho\lambda (\ln \Omega_i)_{,z}}
{\rho^2+\lambda^2}.\nonumber
\end{eqnarray}

If $\hat \psi(\Omega_i)$ and $\hat \psi(\omega_i)$ 
are solutions of Eq. (\ref{eq:con3}),
we can find that $\hat \psi(\Omega_i\omega_i)$
is also the solution of Eq. (\ref{eq:con3}) and satisfies the relation
$\hat \psi(\Omega_i\omega_i)=\hat \psi(\Omega_i)\hat \psi_i(\omega_i)$.
We list the relevant solutions of Eq. (\ref{eq:con3})
in our discussion
\begin{eqnarray}
& &\hat \psi(1)=1, \quad
\hat \psi(\rho^2)=\rho^2-2z\lambda-\lambda^2 
\label{eq:severalpsi}\\
& &\hat \psi (\lambda_1)=\lambda_1-\lambda, \quad
\hat \psi (\lambda_2)=\lambda_2+\lambda,
\end{eqnarray}
where $\lambda_1$ and $\lambda_2$ are given by
\begin{eqnarray}
\lambda_1:&=&\sqrt{\rho^2+(z-z_0)^2}-(z-z_0),
\\
\lambda_2:&=&\sqrt{\rho^2+(z-z_0)^2}+(z-z_0).
\end{eqnarray}
Here $z_0$ is an arbitrary constant.

\section{Generation of seed solution for black ring}
\label{sec:seed}

As have been shown in \cite{Tomizawa},
the black ring solution rotating $S^2$-direction
\cite{Mishima:2005id} is obtained as a two-soliton solution
by taking the five-dimensional Minkowski spacetime
as a seed solution.
In order to obtain the black ring solution
with $S^1$-rotation, we start from the following metric:
\begin{eqnarray}
ds^2=-dt^2+\rho ^{1+d}d\psi ^2+\rho ^{1-d} d\phi ^2
+\rho ^{(d^2-1)/2}(d\rho ^2+dz^2).
\label{eq:levicivita}
\end{eqnarray}
where $d$ is a constant. 
The metric (\ref{eq:levicivita}) is the direct product of 
the Euclidean Levi-Civita metric and one-dimensional flat space.
Thus it is a vacuum solution of five-dimensional Einstein equations.
Performing the Wick rotation $\psi \to i\bar t$,
the four-dimensional base space describes
the Levi-Civita solution
\begin{eqnarray}
ds^2=-\rho ^{1+d}d\bar t^2+\rho ^{1-d} d\phi ^2
+\rho ^{(d^2-1)/2}(d\rho ^2+dz^2).
\label{eq:llevicivita}
\end{eqnarray}
We can find that the Minkowski and Rindler spacetime
are recovered by setting $d=-1, 1$, respectively.
$d=0$ corresponds to one of the Kinnersley's type D
metric \cite{kinnersley}.
From the Levi-Civita metric (\ref{eq:llevicivita}),
we can obtain the Weyl class solutions
in four dimensions by adding solitons \cite{letelier1984}.

For the diagonal seed solutions such as (\ref{eq:levicivita}),
diagonal solutions can be obtained by taking some of the parameters
vanishing, e.g., $m_{01}^{(k)}=m_{02}^{(k)}=0$.
The assumption that the seed solution $g_0$
is of diagonal form quite simplifies analyses
since the generating matrix
$\Psi_0$ may be set to be diagonal
$\Psi_0={\rm diag} (\psi_1,\psi_2,\psi_3)$, where
$\psi_i$ are functions which depend on $\lambda, \rho$ and $z$.
If this is the case, the partial differential equations
for the generating matrix $\Psi_0$ are decoupled
into each component, which allow us to solve each
$\psi_i$ independently.

We should comment on the normalization of the metric
little more carefully. In the previous section,
we have normalized all the metric components
by the same weight as seen in Eq. (\ref{eq:nor}).
Nevertheless, we should note that there exists some freedom
for the normalization when we construct a physical metric 
for the case in which the generated solution has at most
one non-diagonal component.
The main reason is that in this particular case
the soliton equations (\ref{eq:soliton}) are
decomposed into each component.
As we will see, we adopt a distinct normalization
for our solution because some components of the generated
solution is identical to those of the seed
(see Appendix \ref{sec:app2}).
Taking the case of $m_{01}^{(k)}=0$ for example,
the three-dimensional unphysical metric
$g^{{\rm (unphys)}}$ is decomposed into
the $2+1$ block matrix as

\begin{eqnarray}
g^{{\rm (unphys)}}=\left(
\begin{array}{@{\,}c|ccc@{\,}}
(g_0)_{11}  & 0 \\ \hline
0  & g^{{\rm (unphys)}}_{AB}
\label{eq:21}
\end{array}
\right),
\end{eqnarray}
where $g^{{\rm (unphys)}}_{AB} (A,B=2,3)$ is
a $2\times 2$ matrix dependent on the $2n$ parameters
$m_{02}^{(k)},m_{03}^{(k)} (k=1,...,n)$, i.e.,
we add solitons exclusively to the block diagonal
components $(g_0)_{AB}$.
In the present case,
we may choose a normalization which multiplies
only $g^{{\rm (unphys)}}_{AB}$ by the normalization factor
in such a way that the metric satisfies the supplementary
condition ${\rm det}g=-\rho ^2$.
Namely,
we leave the component $(g_0)_{11}$ intact, i.e.,
\begin{eqnarray}
g^{{\rm (phys)}}=\left(
\begin{array}{@{\,}c|ccc@{\,}}
(g_0)_{11}& 0 \\ \hline 0  &
\displaystyle \left(\prod_{k=1}^n\frac{\mu_k}{\rho}\right)
g^{{\rm (unphys)}}_{AB}
\label{eq:norm23}
\end{array}
\right).
\end{eqnarray}
We can easily show that if a seed metric satisfies the condition
${\rm det} g_0=-\rho^2$, the physical metric (\ref{eq:norm23}) also
satisfies this condition.
The general expression for the diagonal $n$-soliton solution
under the normalization (\ref{eq:norm23}) is
\begin{eqnarray}
g_{tt}=(g_0)_{tt}, \qquad
g_{\psi \psi}=\left[\prod _{k=1}^n
\left(\frac{\mu _k}{\rho }\right)\right]
(g_0)_{\psi \psi}, \qquad
g_{\phi \phi}=-\frac{\rho ^2}{g_{tt}g_{\psi \psi}}.
\label{eq:metric}
\end{eqnarray}

Here we focus attention on the three real-pole trajectories
and the Levi-Civita metric with $d=0$ is taken as a seed.
As in \cite{Iguchi}, 
we choose the three parameters $w_k (k=1,2,3)$ such that
\begin{eqnarray}
w_1=-\eta_1\sigma, \qquad w_2=\eta_2\sigma,\qquad
w_3=\kappa\sigma,
\end{eqnarray}
where $\eta_1,\ \eta_2$, $\kappa $ are constants
satisfying
\begin{eqnarray}
\kappa\ge 1, \qquad -1<\eta_1<1,\qquad -1<\eta_2<1,\qquad
\eta_1+\eta_2>0.
\label{eq:parameters}
\end{eqnarray}
%corresponding to Eq. (\ref{eq:c-metric}).
The pole trajectories are given by
\begin{eqnarray}
\mu_1^\pm &=& \pm R_{-\eta_1\sigma }-(z+\eta_1\sigma),\nonumber \\
\mu_2^\pm &=& \pm R_{\eta_2 \sigma }-(z-\eta_2\sigma),\\
\mu_3^\pm &=& \pm R_{\kappa \sigma }-(z-\kappa\sigma),\nonumber
\end{eqnarray}
where we have set
\begin{eqnarray}
R_d:=\sqrt{\rho ^2+(z-d)^2}
\end{eqnarray}
Note that $\mu _k^+\mu_k^- =-\rho ^2$
for each $k$.
The resulting metric component is
\begin{eqnarray}
& & g_{\psi\psi}=\frac{\mu _1^-\mu _2^+\mu _3^-}{\rho ^2}
\label{eq:3soliton1}\\
& &=\frac{(R_{\kappa \sigma }+z-\kappa\sigma)
(R_{\eta _2\sigma }-z+\eta_2\sigma)}
{R_{-\eta _1 \sigma }-z-\eta_1\sigma}.
\label{eq:3soliton}
\end{eqnarray}
The sign choice in Eq. (\ref{eq:3soliton1}) is for
the sake of convenience.
From Eq. (\ref{eq:metric}), we have
\begin{eqnarray}
g_{\phi \phi}=\frac{(R_{-\eta _1\sigma }-z-\eta_1\sigma )
(R_{\kappa \sigma }-z+\kappa\sigma)}
{R_{\eta_2 \sigma }-z+\eta_2\sigma},
\end{eqnarray}
from which we can obtain the rest of the metric components
by the straightforward quadratures (\ref{eq:f1}) and (\ref{eq:f2}).
Making the coordinate transformation in Appendix
\ref{sec:app2},
we can find the resulting three-soliton solution
is the direct product spacetime of the Euclidean vacuum C-metric
\cite{C-metric,Hong2003}
and the one-dimensional flat space
\begin{eqnarray}
ds^2=-dt^2+\frac{2\tilde \kappa ^2}{(u-v)^2}\left[
-G(u)d\psi^2+G(v)d\phi ^2-\frac{du^2}{G(u)}
+\frac{dv^2}{G(v)} \right],
\label{eq:c-metric}
\end{eqnarray}
where $G(\xi )$ is the structure function
given by
\begin{eqnarray}
G(\xi )=(1-\xi ^2)(1+c\xi ).
\end{eqnarray}
Since the metric (\ref{eq:c-metric})
is the Euclidean vacuum C-metric plus
one-dimensional flat space, it
is obviously the vacuum solution of the
five-dimensional Einstein equations.
The C-metric describes a pair of point particles undergoing
uniform acceleration \cite{C-metric}.
$A=(\sqrt{2}\tilde \kappa )^{-1}$ and
$m=c\tilde \kappa /\sqrt{2}$ represents the
acceleration and mass of the point particles, respectively.

In the next section, we use the solution
\begin{eqnarray}
ds^2=-dt^2+g_2d\phi ^2+g_3d\psi ^2,
\label{eq:seed}
\end{eqnarray}
as a seed to obtain the black ring solution,
where

\begin{eqnarray}
g_2:=\frac{(R_{-\eta_1\sigma}-z-\eta_1\sigma)(R_{\kappa\sigma}-z+\kappa\sigma )
}{(R_{\eta_2\sigma}-z+\eta_2\sigma)},
\qquad
g_3:=\frac{(R_{-\eta_1\sigma}+z+\eta_1\sigma)(R_{\eta_2\sigma}-z+\eta_2\sigma)
}{(R_{\kappa\sigma}-z+\kappa\sigma)}.
\label{eq:g2g3}
\end{eqnarray}
The seed solution (\ref{eq:seed}) satisfies
the supplementary condition ${\rm det}g=-\rho ^2$ and coincides 
with the seed obrained by Iguchi and Mishima to generate 
the black ring solution with the $S^1$-rotation~\cite{Iguchi}. 
For $\kappa =\eta_2$ case, the seed (\ref{eq:seed})
reduces to the Minkowski spacetime.
Supposed $\Psi={\rm diag}(\psi_1,\psi_2,\psi_2)$,
we can find from the analysis in Sec.\ref{sec:rescale}
that the generating matrix for the background metric (\ref{eq:seed}) is expressed as
\begin{eqnarray}
& &\psi_1[\lambda ]=-1,\nonumber \\
& &\psi_2[\lambda ]=\frac{(R_{-\eta_1\sigma}-z-\eta_1\sigma-\lambda)(R_{\kappa\sigma}-z+\kappa\sigma-\lambda)
}{ (R_{\eta_2\sigma}-z+\eta_2\sigma-\lambda )},
\label{eq:generating}\\
& &\psi_3[\lambda ]=
\frac{(R_{-\eta_1\sigma}+z+\eta_1\sigma+\lambda )(R_{\eta_2\sigma}-z+\eta_2\sigma-\lambda )}{(R_{\kappa\sigma}-z+\kappa\sigma-\lambda )},\nonumber
\end{eqnarray}
where we have omitted the argument $\rho$ anz $z$ 
in each $\psi _i$ for simplicity,
which is understood to represent 
$\psi_i[\lambda]=\psi_i(\lambda, \rho, z)$.

\section{Production of black ring solution}
\label{sec:black ring}

In view of the fact that the $\psi-\psi$ component
of the seed solution coincides that of the
black ring of Emparan and Reall
(see Appendix \ref{sec:app2}),
it is suitable to normailze the metric so that
$(g_0)_{33}$ is unchanged:
\begin{eqnarray}
g^{{\rm (phys)}}=\left(
\begin{array}{@{\,}c|ccc@{\,}}
\displaystyle \left(\prod_{k=1}^n
\frac{\mu_k}{\rho}\right)
g^{{\rm (unphys)}}_{AB}
& 0 \\ \hline 0  &
(g_0)_{33}
\label{eq:norm12}
\end{array}
\right),
\end{eqnarray}
where $A, B=1,2$.
Here, we consider the two-soliton solution.
 We choose the sign of plus in Eq. (\ref{eq:mup})
and take the constants $w_1=-w_2=-\sigma $
without loss of generality.
Under the special normalization (\ref{eq:norm12}),
the two-soliton solution can be written 
in the following form:

\begin{eqnarray}
& &g^{{\rm (phys)}}_{tt}=
-\frac{G_{tt}}{\mu_1\mu_2\Sigma},\quad
g_{t\phi}^{{\rm (phys)}}
=-g_2\frac{(\rho^2+\mu_1\mu_2)
G_{t\phi}}{\mu_1\mu_2 \Sigma},\quad
g^{{\rm (phys)}}_{\phi\phi}=
-g_2\frac{G_{\phi\phi}}{\mu_1\mu_2\Sigma},
\label{eq:gphys}
\\
& &g^{{\rm (phys)}}_{\psi\psi}=g_3,\quad
g_{\phi\psi}^{{\rm (phys)}}=g_{t\psi}^{{\rm (phys)}}=0,
\end{eqnarray}
where the functions 
$G_{tt},\ G_{t\phi},\ G_{\phi\phi}$
and $\Sigma$ are defined as
\begin{eqnarray}
& &G_{tt}=-m_{01}^{(1)2}m_{01}^{(2)2}
\psi_2[\mu_1]^2\psi_2[\mu_2]^2(\mu_1-\mu_2)^2
\rho^4+m_{01}^{(1)2}m_{02}^{(2)2}g_2\mu_2^2
(\rho^2+\mu_1\mu_2)^2\psi_2[\mu_1]^2
\nonumber\\
& &+m_{01}^{(2)2}m_{02}^{(1)2}g_2 \mu_1^2
(\rho^2+\mu_1\mu_2)^2\psi_2[\mu_2]^2
-m_{02}^{(1)2}m_{02}^{(2)2}g_2^2
\mu_1^2\mu_2^2(\mu_1-\mu_2)^2\\
& &-2m_{01}^{(1)}m_{01}^{(2)}m_{02}^{(1)}m_{02}^{(2)}
g_2\psi_2[\mu_1]\psi_2[\mu_2](\rho^2+\mu_1^2)
(\rho^2+\mu_2^2)\mu_1\mu_2,
\nonumber \\\nonumber \\
& &G_{\phi\phi}=m_{01}^{(1)2}m_{01}^{(2)2}
\mu_1^2\mu_2^2(\mu_1-\mu_2)^2
\psi_2[\mu_1]^2\psi_2[\mu_2]^2
+m_{02}^{(1)2}m_{02}^{(2)2}g_2^2
(\mu_1-\mu_2)^2\rho^4\nonumber\\
& &-m_{01}^{(1)2}m_{02}^{(2)2}g_2\mu_1^2
\psi_2[\mu_1]^2(\rho^2+\mu_1\mu_2)^2
-m_{01}^{(2)2}m_{02}^{(1)2}g_2\mu_2^2
(g_2-\mu_2)^2(\rho^2+\mu_1\mu_2)^2  \\
& &+2m_{01}^{(1)}m_{01}^{(2)}m_{02}^{(1)}m_{02}^{(2)}
g_2\mu_1\mu_2\psi_2[\mu_2]\psi_2[\mu_1]
(\rho^2+\mu_1^2)(\rho^2+\mu_2^2),
\nonumber \\ \nonumber \\
& &G_{t\phi}=m_{01}^{(1)}m_{01}^{(2)2}
m_{02}^{(1)}\mu_2(\mu_1-\mu_2)
\psi_2[\mu_2]^2\psi_2[\mu_1](\rho^2+\mu_1^2)
+m_{01}^{(1)}m_{02}^{(1)}m_{02}^{(2)2}
g_2\mu_2(\mu_2-\mu_1)
\psi_2[\mu_1](\rho^2+\mu_1^2)\nonumber
\\
& &+m_{01}^{(1)2}m_{01}^{(2)}m_{02}^{(2)}
\mu_1(\mu_2-\mu_1)\psi_2[\mu_1]^2
\psi_2[\mu_2](\rho^2+\mu_2^2)
+m_{01}^{(2)}m_{02}^{(1)2}m_{02}^{(2)}
\mu_1g_2\psi_2[\mu_2](\rho^2+\mu_2^2)(\mu_1-\mu_2),
\\ \nonumber \\
& &\Sigma=m_{01}^{(1)2}m_{01}^{(2)2}
\psi_2[\mu_1]^2\psi_2[\mu_2]^2(\mu_1-\mu_2)^2\rho^2
+m_{02}^{(1)2}m_{02}^{(2)2}g_2^2(\mu_1-\mu_2)^2\rho^2
+m_{01}^{(1)2}m_{02}^{(2)2}g_2\psi_2[\mu_1]^2
(\rho^2+\mu_1\mu_2)^2\nonumber\\
& &+m_{02}^{(1)2}m_{01}^{(2)2}g_2
\psi_2[\mu_2]^2(\rho^2+\mu_1\mu_2)^2
-2m_{01}^{(1)}m_{01}^{(2)}m_{02}^{(1)}m_{02}^{(2)}
g_2\psi_2[\mu_1]\psi_2[\mu_2](\rho^2+\mu_1^2)(\rho^2+\mu_2^2),
\end{eqnarray}
where the two functions $g_2$ and $g_3$ are given by Eq. (\ref{eq:g2g3}).

In order for the metric to approach the Minkowski spacetime
asymptotically,
let us consider the coordinate transformation
of the physical metric such that
\begin{eqnarray}
t\rightarrow t'=t-C_1\phi, \qquad \phi
\rightarrow \phi'=\phi,
\end{eqnarray}
where $C_1$ is a constant chosen to ensure the
asymptotic flatness.
We should note that the transformed metric also satisfies the
supplementary condition ${\rm det} g=-\rho^2$.
Under this transformation, the physical metric components become
\begin{eqnarray}
& &g_{tt}^{\rm (phys)}\rightarrow
g_{t't'}^{\rm (phys)}=g_{tt}^{\rm (phys)},
\nonumber \\
& &g_{t\phi}^{\rm (phys)}\rightarrow
g_{t'\phi'}^{\rm (phys)}=g_{t\phi}^{\rm (phys)}+C_1
g_{tt}^{\rm (phys)},\\
& &g_{\phi\phi}^{\rm (phys)}\rightarrow
g_{\phi'\phi'}^{\rm (phys)}=g_{\phi\phi}^{\rm (phys)}
+2C_1 g_{t\phi}^{\rm (phys)}+C_1^2g_{tt}^{\rm (phys)}.
\nonumber
\end{eqnarray}
If we choose the parameters such that
\begin{eqnarray}
m_{01}^{(1)}m_{01}^{(2)}&=&\beta,\\
m_{01}^{(2)}m_{02}^{(1)}&=&\sigma^{\frac{1}{2}}(\kappa_2-1),\\
m_{01}^{(1)}m_{02}^{(2)}&=&-\sigma^{\frac{1}{2}}\alpha\beta(\kappa_1+1),\\
m_{02}^{(1)}m_{02}^{(2)}&=&-\sigma\alpha(\kappa_1+1)(\kappa_2-1), \\
C_1 &=&\frac{2\sigma ^{1/2}\alpha }{1+\alpha \beta },
\end{eqnarray}
with
\begin{eqnarray}
\kappa_1+1=\frac{(-\kappa+1)(\eta_1+1)}{-\eta_2+1},
\qquad \kappa_2-1=\frac{(\kappa+1)(\eta_1-1)}{\eta_2+1},
\end{eqnarray}
and use the prolate spheroidal coordinate $(x,y)$ defined by
\begin{eqnarray}
\rho=\sigma\sqrt{(x^2-1)(1-y^2)},\qquad z=\sigma xy,
\label{eq:prolate}
\end{eqnarray}
we can confirm
that the transformed metric coincides
with the expression of a black ring solution
in \cite{Iguchi} with the aid of the equations
in Appendix.{\ref{sec:app1}};
\begin{eqnarray}
ds^2=-\frac{A}{B}\left[dt-\left(2\sigma g_3^{-1/2}
\frac{C}{A}
+C_1\right)d\phi \right]^2
+\frac{B}{A}g_3^{-1}\rho ^2d\phi ^2+g_3d\psi ^2
+C_2 \frac{x^2-y^2}{x^2-1}Be^{2\gamma'}g_3^{-1}
\left(\frac{dx^2}{x^2-1}+\frac{dy^2}{1-y^2}\right),
\label{eq:blackring}
\end{eqnarray}
where
\begin{eqnarray}
& &A:=(x^2-1)(1+FG)^2-(1-y^2)(F-G)^2,
\label{eq:A}\\
& &B:=[(x+1)+(x-1)FG]^2+[(1+y)F+(1-y)G]^2,
\label{eq:B}\\
& &C:=(x^2-1)(1+FG)[G-F-y(F+G)]
+(1-y^2)(G-F)[1+FG+x(1-FG)],
\label{eq:C}\\
& &F:=-\alpha\frac{\sigma^{3/2}(-\kappa+1)(\eta_1+1)}
{-\eta_2+1}\frac{g_2^{1/2}(x+1)(1-y)}{\psi_2[\mu_2]\rho},
\label{eq:F}\\
& &G:=-\beta\frac{(\eta_2+1) }{(\kappa+1)(\eta_1-1)
\sigma^{3/2}}\frac{\rho \psi_2[\mu_1]}{g_2^{1/2}(x-1)(1-y)},
\label{eq:G}\\
& &\gamma':=\frac{1}{4}\left[\ln \left(
U_\sigma U_{-\sigma }^{-1}
U_{\kappa \sigma }^2U_{-\eta_1\sigma }^2
U_{\eta_2\sigma }^{-2}
\right)\right.
\nonumber \\
& &\left. ~~~~
-\ln \left(Y_{\sigma ,\sigma }
Y_{-\sigma ,-\sigma }
Y_{\kappa \sigma ,\kappa \sigma }
Y_{-\eta _1\sigma ,-\eta_1\sigma }
Y_{\eta_2\sigma \eta_2\sigma }
Y_{\sigma ,-\sigma }^{-2}
Y_{\sigma ,\kappa \sigma }
Y_{-\sigma ,\kappa \sigma }^{-1}
\right.\right. \\ \label{eq:f'}
& &\left.\left.\times
Y_{\sigma ,-\eta_1\sigma }
Y_{\sigma ,\eta_2\sigma }^{-1}
Y_{-\sigma ,-\eta_1\sigma }^{-1}
Y_{-\sigma ,\eta_2\sigma }
Y_{\kappa \sigma ,-\eta_1\sigma }^2
Y_{\kappa \sigma ,\eta_2\sigma }^{-2}
Y_{\eta_1\sigma ,\eta_2\sigma }^{-2}
\right)\right],
\nonumber \\
& &U_c:=R_c+z-c, \\
& &Y_{c,d}:=R_cR_d+(z-c)(z-d)+\rho ^2,\\
& &C_2:=\frac{\sigma ^2}{(1+\alpha \beta)^2 }.
\end{eqnarray}
The constant $C_2$ is chosen to assure 
the asymptotic flatness of the solution.
When $\kappa =\eta_2$,
this solution reduces to the one obtained
in \cite{Tomizawa},
which is the $S^2$-rotating black ring solution.
At first sight, the solution (\ref{eq:blackring})
includes six parameters
$\sigma ,\alpha, \beta, \kappa ,\eta _1, \eta_2 $,
but one of which is a kinematical parameter so that
one can freely choose the value of it.
The solution is then specified by the remaining five parameters.
This solution, however, has closed timelike curves in general.
The condition for the absence of closed timelike curves
restricts the value of $\alpha $ and $\beta $ as
\cite{Iguchi}
\begin{eqnarray}
\alpha =\sqrt{\frac{2(1-\eta_2)}{(\kappa -1)(1+\eta_1)}},
\qquad
\beta =\sqrt{\frac{(\kappa +1)(1-\eta _1)}{2(1+\eta_2)}}.
\label{eq:noctc}
\end{eqnarray}
These follow from the condition that the orbit
$(\partial /\partial \phi)$ should close at $(x,y)=(1,\pm 1)$.
The remaining free parameters are three.
However, the above condition is not sufficient for
the metric to be well-posed throughout the spacetime
since the solution still has conical singularities.
The lack of conical singularities, furthermore,
imposes the following restriction on the rest of three
parameters to
\begin{eqnarray}
1+\alpha \beta =\sqrt{\frac{\kappa +1}{\kappa -1}}
\left(\frac{\kappa-\eta _2}{\kappa +\eta_1}
\right)\left(1+\alpha \beta \frac{\kappa -1}{\kappa +1}
\right).
\label{eq:noconical}
\end{eqnarray}
Consequently, the resultant metric is characterized
by two parameters. %, that is, the mass and angular momentum.
As shown in Appendix \ref{sec:app2}, under these physically
reasonable conditions, this solution (\ref{eq:blackring})
accords with the one found by Emparan and Reall.
Thus, the remaining two parameters are concerned with the mass
and angular momentum, which specifies the black ring solution.

\section{Summary and discussion}
\label{sec:conclusion}

In this article, we studied the inverse scattering method
for the asymptotically flat, stationary and axisymmetric spacetimes
of the five-dimensional vacuum Einstein equations.
Our main result is the reproduction of the black ring solution
found by Emparan and Reall via the inverse scattering technique.
The steps of the procedure we have done is as follows:
in the first place, we consider the direct product spacetime
with one-dimensional timelike flat space and a four-dimensional
base space. The four-dimensional base space is taken to be
the Euclidean Levi-Civita solution with $d=0$
(\ref{eq:llevicivita}), whence
this metric is  apparently the vacuum solution of the five-dimensional
Einstein equations with cylindrical symmetry.
Considering the fact that the static black ring solution
belongs to the five dimensional Weyl class solutions,
it may be reasonable to take the Levi-Civita metric
as a starting point since it describes the family of
the cylindrically symmetric static solutions of vacuum
Einstein equations,
from which we can obtain the Weyl class solutions in four dimensions
by adding solitons via the inverse scattering method.
Subsequently, the Euclidean vacuum C-metric is generated as a three-soliton
solution from the Levi-Civita background with $d=0$ \cite{letelier1984}.
The appearance of the C-metric seems to be convincing
since the original derivation of black ring solution have
employed the Euclidean dilatonic C-metric.
Lastly, the black ring solution with $S^1$-rotation
is constructed as a two-soliton solution
from the seed of the C-metric plus
one-dimensional flat space.
This solution coincides with the one obtained by
Iguchi and Mishima \cite{Iguchi}, and
encompasses broader class than the one
found by Emparan and Reall. But it generally
has closed timelike curves and conical singularity 
in the vicinity of the event horizon.
If we impose that the solution is free from causal violation,
we finally obtain the black ring solution of Emparan and Reall
(see Appendix \ref{sec:app2}). 
The condition for the absence of conical singularity 
comes down to the completely regular black ring solution,
which is characterized by the mass and angular momentum.
Our consequence is the first example that clarifies the
physical origin of the seed solution and indicates the possibility
of finding new exact solutions in higher dimensions.

The point we would like to stress is the special normalization
such as (\ref{eq:norm23}) and (\ref{eq:norm12}), namely
some of the components of the seed remain unchanged.
Even the five-dimensional Schwarzschild solution is
generated in this way \cite{Koikawa}.
We also comment that if we want to
obtain a non-trivial regular black holes/ring solution 
with non-vanishing two angular momentum components,
the overall normalization (\ref{eq:nor}) does not seem to 
yield regular spacetimes as two-soliton solutions
unless the seed solution satisfies the 
condition ${\rm det}g=-\rho ^2$.
To obtain the metric with non-vanishing off-diagonal
components, it may be ineluctable to adopt
some technical seed in the sense that it does not 
satisfy the condition ${\rm det}g=-\rho ^2$, 
which is to say, 
it is not the solution of vacuum Einstein equations
although it is the solutions of Lax pair. 
In fact, the seed to generate the Myers-Perry 
black hole solution with two non-vanishing angular momentum 
components in \cite{Pomeransky:2005sj} does not meet
${\rm det}g=-\rho ^2$.  
Otherwise, it may be necessary to evaluate the
multi-soliton solutions.

It is also worth noting that
the generated solution (\ref{eq:blackring})
reduces to the $S^2$-rotating black ring solution
when $\kappa =\eta_2$ with $\eta_1>1$,
because the Minkowski spacetime is recovered when
$\kappa =\eta_2$ in the C-metric seed
(see Eqs. (\ref{eq:seed}) and (\ref{eq:g2g3})). 
In consequence, our generation is the most general one
rather than those heretofore discussed 
in that all the previous results are recaptured
as special cases of our model;
the $S^2$-rotating black ring solution \cite{Tomizawa},
the Myers-Perry solution with one angular momentum
\cite{Azuma} and of course the five-dimensional
Schwarzschild solution \cite{Koikawa} are realized.

Complete integrability of the Einstein equations
highly resorts to the fact that the metric is written by
the canonical form (\ref{eq:canonical}) on top of that
we focus our attention exclusively on the solitonic solutions.
In higher dimensions than five, even the Minkowski spacetime
does not fall into this type due to the lack of
Killing vectors meeting certain requirements.
Accordingly our discussions cannot
straightforwardly applicable into higher dimensions than five.
In order to obtain such solutions, 
we might have to formulate the inverse scattering method 
in other coordinate system rather than the canonical coordinate
(\ref{eq:canonical}).

As mentioned in the introduction,
all asymptotically flat five-dimensional vacuum black hole solutions 
with a connected component of the event horizon
which has already been discovered in this stage can be produced
via the inverse scattering method.
We expect to be able to construct new higher-dimensional
exact solutions by this powerful tool
although we have not yet found such solutions.
We can speculate from the results in the present paper
that an infinite class of higher-dimensional
stationary and axisymmetric solutions of vacuum 
Einstein equations 
can be generated by means of the inverse scattering method
by taking the higher-dimensional Weyl class solution as a seed.
Besides that, it may also be of interest to include several matter fields 
appearing in string theory and supergravity.
These prospective issues are worth investigating
in the future.

\section*{Acknowledgements}
We would like to thank Takashi~Mishima and Hideo~Iguchi
for continuous discussion. We also thank Ken-ichi~Nakao and
Hideki~Ishihara for continuous encouragement.
MN would like to thank Kei-ichi Maeda for a lot of
fruitful discussions and Yukinori Yasui for
useful comments. We are also thankful to Edward Teo for valuable comments. 
MN is partially supported by the Waseda University Grant
for The 21st Century COE Program
(Holistic Research and Education Center for Physics
Self-organization Systems) at Waseda University.

\appendix

\section{C-metric coordinates and canonical form}
\label{sec:app2}

Define the C-metric coordinates $(u, v)$ as
\begin{eqnarray}
\rho :=\frac{2\tilde \kappa ^2\sqrt{-G(u)G(v)}}{(u-v)^2}, \qquad
\tilde{z}:=z+\frac{\eta_1-\eta_2}{2}\sigma =
\frac{\tilde\kappa ^2(1-uv)(2+cu+cv)}{(u-v)^2},
\end{eqnarray}
where $G(\xi )$ is a structure function given by
\begin{eqnarray}
G(\xi )=(1-\xi ^2)(1+c \xi).
\end{eqnarray}
We have introduced new constants $c~(0<c<1)$
and $\tilde{\kappa }$ defined as
\begin{eqnarray}
c&=&\frac{\eta_1+\eta_2}{2\kappa +\eta_1-\eta_2},
\label{eq:c}\\
\tilde{\kappa }^2&=&\left(\kappa +\frac{\eta_1-\eta_2}{2}
\right)\sigma.
\label{eq:kappa}
\end{eqnarray}
The coordinate $u$ and $v$ lie in the ranges
\begin{eqnarray}
-1\leq u \leq 1, \qquad
v\leq -1.
\end{eqnarray}
Defining a new constant $b$ as
\begin{eqnarray}
b=\frac{[\kappa +1+(\kappa -1)\alpha \beta]^2
-(\kappa ^2-1)(1+\alpha \beta )^2 }
{[\kappa +1+(\kappa -1)\alpha \beta]^2
+(\kappa ^2-1)(1+\alpha \beta )^2},
\label{eq:b}
\end{eqnarray}
the black ring solution (\ref{eq:blackring})
with the conditions (\ref{eq:noctc}) and
(\ref{eq:noconical}) is then transcribed as
\cite{Emparan:2001wn,Harmark:2004rm}
\begin{eqnarray}
ds^2=-\frac{F(v)}{F(u)}\left(dt
-\mathcal C\kappa \frac{1+v}{F(v)}d\phi \right)^2
+\frac{2\tilde \kappa ^2F(u)}{(u-v)^2}
\left[-\frac{G(v)}{F(v)}d\phi ^2
+\frac{G(u)}{F(u)}d\psi ^2+
\frac{du^2}{G(u)}-\frac{dv^2}{G(v)}
\right]
\label{eq:blackring2}
\end{eqnarray}
where
\begin{eqnarray}
F(\xi )&=&1+b\xi , \\ %\qquad
\mathcal C&=&\sqrt{2b(b-c)\frac{1+b}{1-b}}.
\end{eqnarray}
The parameters $b$ and $c$ take the value
in the range
\begin{eqnarray}
0<c\leq b<1.
\end{eqnarray}
The ring (\ref{eq:blackring2}) generally has the
conical singularity, which is cured by choosing 
the parameter $b$ such that
\begin{eqnarray}
b=\frac{2c}{1+c^2}.
\end{eqnarray}
In the canonical coordinate $\rho $ and $z$,
the metric is rewritten as
\begin{eqnarray}
& &g_{11}=-\frac{(1+b)(1-c)R_1+(1-b)(1+c)R_2-2(b-c)R_3-2b
(1-c^2)\tilde{\kappa }^2}{(1+b)(1-c)R_1+(1-b)(1+c)R_2-2(b-c)R_3+2b
(1-c^2)\tilde{\kappa }^2},\\
& &g_{12}=-\frac{2\mathcal C\tilde{\kappa }
(1-c)[R_3-R_1+(1+c)\tilde{\kappa }^2]}
{(1+b)(1-c)R_1+(1-b)(1+c)R_2-2(b-c)R_3+2b
(1-c^2)\tilde{\kappa }^2},\\
& &g_{22}=-\frac{\rho ^2}{g_{11}g_{33}}+\frac{g_{12}^2}{g_{11}},\\
& &g_{33}=%\frac{(R_1+R_2+2c\tilde{\kappa }^2)
\frac{(R_3+z-\tilde\kappa ^2)(R_2-z+c\tilde \kappa ^2)}
{R_1-z-c\tilde{\kappa }^2},\label{eq:g33ring}\\
& &f=[(1+b)(1-c)R_1+(1-b)(1+c)R_2-2(b-c)R_3
+2b(1-c^2)\tilde{\kappa}^2]\times
\frac{(1-c)R_1+(1+c)R_2+2cR_3}{8(1-c^2)^2R_1R_2R_3},
\end{eqnarray}
where we have defined
\begin{eqnarray}
R_1=\sqrt{\rho ^2+(\tilde{z}+c\tilde{\kappa }^2)^2},\qquad
R_2=\sqrt{\rho ^2+(\tilde{z}-c\tilde{\kappa }^2)^2},\qquad
R_3=\sqrt{\rho ^2+(\tilde{z}-\tilde{\kappa }^2)^2}.
\end{eqnarray}
Observe that the $g_{33}$ components of the black ring
(\ref{eq:g33ring}) is the same as that of the seed
(\ref{eq:seed}).

\section{Alternative expression for black ring}
\label{sec:app1}

We present the useful expression
for the metric components of a general black ring solution.
Substituting Eqs. (\ref{eq:F}) and (\ref{eq:G}) into
(\ref{eq:A}), (\ref{eq:B}) and (\ref{eq:C}),
another expression of the metric components is derived as
\begin{eqnarray}
g_{tt}=-\frac{\tilde A}{\tilde B}, \quad
g_{t\phi}=2\sigma^{\frac{1}{2}}g_2\frac{\tilde C}{\tilde B}
+C_1\frac{\tilde A}{\tilde B}.\label{eq:MI}
\end{eqnarray}
Here we have introduced new functions
$\tilde A,\ \tilde B$ and $\tilde C$ defined as
\begin{eqnarray}
& &\tilde A=-\beta^2\psi_2[\mu_1]^2\psi_2[\mu_2]^2(1+y)^2
+\sigma\alpha^2\beta^2(\kappa_1+1)^2g_2\psi_2[\mu_1]^2(x+1)^2
+\sigma(\kappa_2-1)^2g_2\psi_2[\mu_2]^2(x-1)^2 \nonumber
\\
& &-\sigma^2\alpha^2(\kappa_1+1)^2(\kappa_2-1)^2g_2^2(1-y)^2
+2\sigma\alpha\beta(\kappa_1+1)(\kappa_2-1)g_2\psi_2[\mu_1]
\psi_2[\mu_2](x^2-y^2),
\\ \nonumber \\
& &\tilde B=\beta^2\psi_2[ \mu_1]^2\psi_2[\mu_2]^2(1-y)^2
+\sigma^2\alpha^2(\kappa_1+1)^2(\kappa_2-1)^2g_2^2(1-y^2)
+\sigma\alpha^2\beta^2(\kappa_1+1)^2\psi_2[\mu_1]^2g_2(x^2-1)
\nonumber
\\
& &+\sigma(\kappa_2-1)^2\psi_2[\mu_2]^2g_2(x^2-1)
+2\sigma\alpha\beta(\kappa_1+1)(\kappa_2-1)g_2
\psi_2[\mu_1]\psi_2[\mu_2](x^2-y^2),
\\ \nonumber \\
& &\tilde C=-\beta(\kappa_2-1)
\psi_2[\mu_1]\psi_2[ \mu_2]^2(x+y)
-\alpha\beta^2(\kappa_1+1)\psi_2[\mu_1]^2
\psi_2[\mu_2](x-y)
\nonumber
\\
& &+\sigma\alpha^2\beta(\kappa_2-1)(\kappa_1+1)^2g_2
\psi_2[\mu_1](x+y)+\sigma\alpha(\kappa_1+1)
(\kappa_2-1)^2\psi_2[\mu_2](x-y),
\end{eqnarray}
where $\psi _2[\mu_1]$ and $\psi_2[\mu_2]$ are obtained
by Eq. (\ref{eq:generating}).

\end{document}